\documentclass[reprint,  amsmath, amssymb, aps, prb]{revtex4-1}

\usepackage{graphicx}
\usepackage{dcolumn}
\usepackage{bm}
\usepackage{color}

\begin{document}

\preprint{APS/123-QED}

\setlength{\abovecaptionskip}{-60pt}

\title{Magneto Acoustic Quantum Oscillations in High Fields and the Fermi Surface of UPt$_3$ }

\author{Ludwig Holleis, V.W. Ulrich and B.S. Shivaram}

\affiliation{Department of Physics, University of Virginia, Charlottesville, VA. 22904 \\}

\date{\today}

\begin{abstract}

We report magneto-acoustic quantum oscillations (MAQO) in the heavy fermion system UPt$_3$ in magnetic fields B, upto 33 T. For B in the ab-plane of the hexagonal crystal MAQO in the sound velocity commence at $\approx$ 12 T and grow with field. However, in contrast to typical Lifshitz-Kosevich behaviour the frequency corresponding to the dominant oscillation increases continuously as the metamagnetic transition (MMT) at 20 T is reached. This dominant MAQO arises from the $\delta$ orbit of band 1 with a large effective mass of 33 m$_e$, for B $<$ 20 T and disappears after the MMT. Thus, the MMT involves a significant change of the Fermi surface, primarily in band 1. For B k ab-plane and $<$ 20 T we reproduce successfully orbits established through previous de Haas-van Alphen and Shubnikov de Haas measurements. We also observe several new orbits, some that can be identified with existing band theory and others not seen previously with completely new frequencies. For B $\parallel$ c-axis we observe MAQOs which also commence at $\approx$ 12 T and grow gradually but break suddenly into a large amplitude and change in frequency at 24.8 T. These enhanced oscillations get weaker again at 30 T. These abrupt changes at 24.8 T and 30 T are signatures of Lifshitz transitions and coincide with the newly discovered spin density wave states for this orientation reported by us recently.

\begin{description}

\item[PACS numbers]
 
75.30.Mb, 71.27.+a, 75.25.Dk

\end{description}

\end{abstract}

\pacs{Valid PACS appear here}

\maketitle

\textbf{I. Introduction:}
The occurance of quantum oscillations (QO) in transport and thermodynamic observables in high magnetic fields is a defining property of a metal. The study of such oscillations in metals together with the successes of Fermi liquid theory have been instrumental in our understanding of the quantum behavior of the electrons in metallic solids and their interpretation in terms of quasiparticles \cite{Abrikosov2017}.  In recent decades, in the context of high temperature superconductors, and in recent years, in the context of topological insulators, there has been a resurgence of interest in employing quantum oscillations to understand these materials\cite{LeyraudNat2007, HartsteinNatPhys2017, GrubinskasPRB2018}.  Their study is thus at the forefront in attempts to precisely define the fundamental behavior of electrons due to many body interactions and the excitations associated with them.  On a more practical note, perhaps due to the relative ease of their implementation in high fields, observations of oscillations in the resistance and the magnetization dominate experimental efforts. Quantum oscillations in other quantities such as thermopower, heat capacity etc. should also occur but are less often pursued.  Another less exploited physical parameter that offers distinct advantages due to the unprecedented precision with which it can be measured is velocity of ultrasound.  Magneto-acoustic quantum oscillations (MAQO) in the longitudinal ultrasound (US) velocity offer the advantage that an extra degree of sensitivity is added when the Fermi surface (FS) is strongly volume dependent \cite{Shoenberg1984, TestardiPR1970}.  This is the case for highly correlated systems such as heavy fermions where the Gruneisen parameter is very large.  The strain dependence is particularly pronounced when small cross sections of the FS and their reconstruction are involved. It is well known that the amplitude of quantum oscillations, in general, is strongly dependent on sample purity and if very large effective masses of the electrons are involved demand the lowest possible temperatures. These requirements are uniform across both small and large orbits on the FS.  However, with US velocity measurements since the observed signals are enhanced due to the strain dependence of the FS, small orbits present a larger fractional change and hence provide more vital information.  Thus, especially small orbits with large effective masses and their influence on overall thermodynamics of strongly correlated systems can be effectively studied.\\

\begin{figure*}
\includegraphics[width=0.85\textwidth]{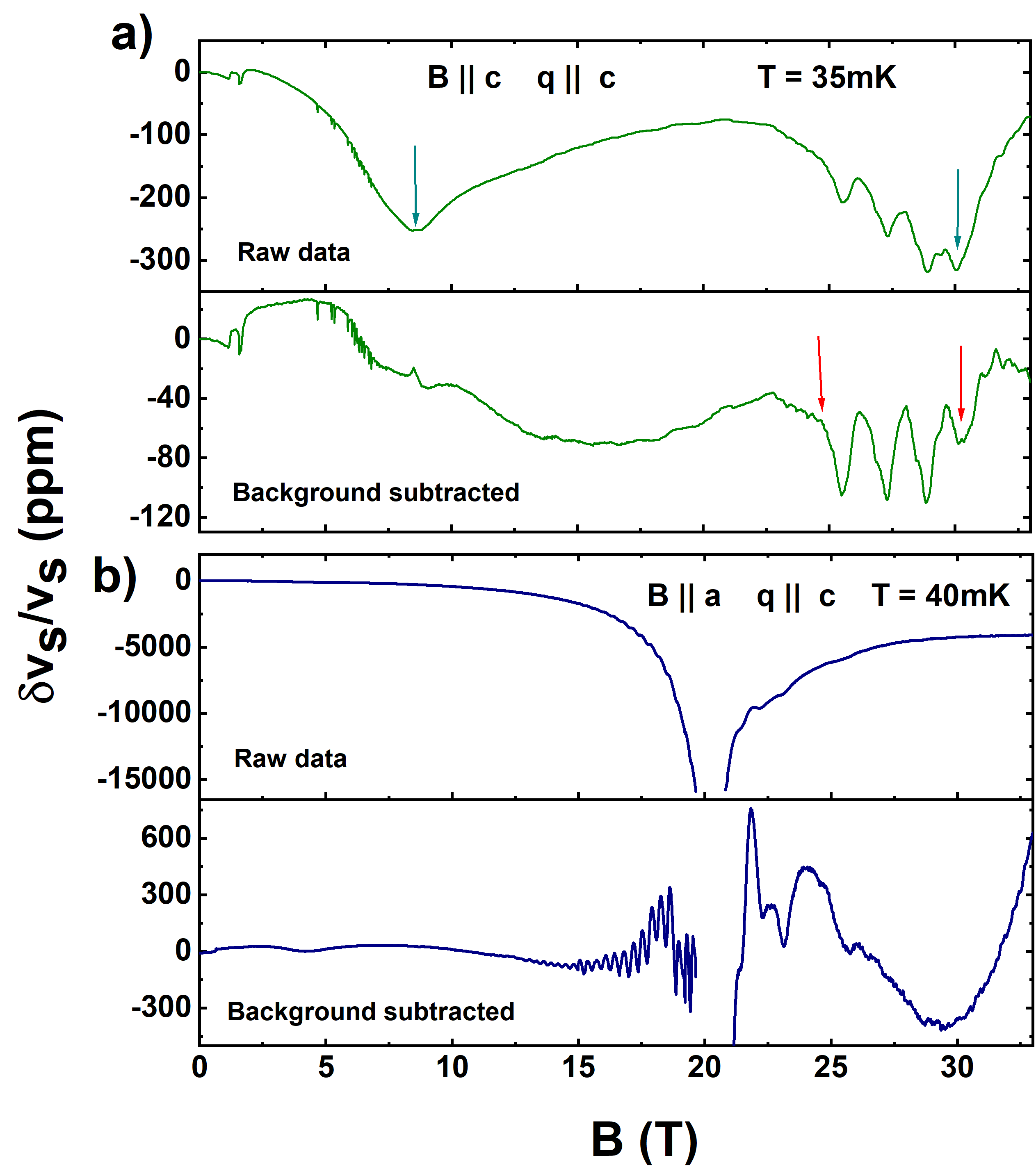}
\vspace{2.2cm}
\caption{\label{fig1}Ultrasound data are shown as measured (top) and with a subtracted background (bottom) along the c- and a-axis on the top and bottom respectively. MAQO are observed along both directions. Along the c-axis several features are observed. The blue arrows indicate dips in the sound velocity corresponding to spin density wave transitions, the red arrows indicate change of the large MAQO assigned to LTs.}
\end{figure*}

\textbf{II. Experimental:} In this communication we report US measurements in UPt$_3$ where we observe many new orbits on the FS predicted by theory \cite{NormanSSC1988, Kimura2000, McMullan2008} but not previously seen in experiments as well as MAQO at new frequencies.  We also observe the orbits seen in previous dHVa and Shubnikov de Haas measurements \cite{Julian1992, Kimura1995, McMullan2008}. In addition, a change in the FS related to the metamagnetic transition in UPt$_3$ at 20 T along the a-axis is seen as well as changes in frequency across the Lifshitz transitions at 24.8 T and 30 T along the c-axis discovered by us recently\cite{ShivaramSR2018}. The measurements we report were performed on zone refined single crystals of approximately 3 mm x 3mm x 2.5 mm grown at Argonne National Labs and of the same quality as those employed in previous ultrasound work \cite{BoukhnyPRB1994}.  A set of four overtone polished quartz transducers, with fundamental resonant frequency of 20 MHz, were glued to the four highly polished crystal faces such that compressional sound waves could be launched with the wavevector either parallel or perpendicular to the hexagonal c-axis.  The UPt$_3$ crystal was subsequently mounted in a specially designed sample holder and the center conductors of four flexible co-axes were contacted with the four transducers.  The sample holder could be fit in the annular space of a brass rotator whose angular position could be set without bringing the sample to room temperature. The position of the rotator and thus the crystal axes with respect to the magnetic field was independently determined with a Hall sensor. A majority of the measurements were performed at the millikelvin facility attached to a 33 T Bitter system at the National High Magnetic Field Laboratory in Tallahassee. Some measurements were also performed in a superconducting magnet in fields upto 20 T. A frequency modulated CW technique was utilized to record shifts in the standing wave resonances (which in turn yield the change in the sound velocity) as the magnetic field was swept at constant temperatures.  \\

\textbf{III. Results:} In figure 1 (a), we show the results of experiments carried out at 35 mK with the field along the c-axis. The top panel shows the raw data where two dips in the sound velocity are apparent. The low field dip at 8 T was assigned by Bruls et al. \cite{Bruls1996} to a spin-density wave (SDW) transition. The high field dip at 30 T recently discovered by us can similarly be attributed as a transition to another SDW state\cite{ShivaramSR2018}. As discussed in that work a separate background subtraction procedure is employed to make the MAQO stand out for this orientation. A large, low frequency MAQO becomes apparent at 24.8 T and disappears again at 30T. We ascribe the signatures at these two fields to Lifshitz Transitions (LT) i.e. a situation where a band is filled or emptied as a function of field. In fig. 1 (b) top panel, the as measured changes in the US velocity at 35 mK with a field up to 33 T applied along the a-axis are shown. The MM transition at 20 T is clearly seen as a large downturn with a subsequent loss of signal due to the large attenuation of sound at the transition. These results are similar to those reported by Feller et al previously\cite{FellerPRB2000}. To make the oscillations stand out we subtract an asymmetric Lorentzian background due to the MM transition and the results are shown in the left bottom panel \cite{Note1}. Distinct and rapid MAQO can clearly be seen below the MM critical field, but these dissappear above the transition. We analyze the various features and the changes in the MAQO more quantitatively through spectral analysis and discuss them below.\\
\\

\begin{figure}[h]
\includegraphics[width=0.49\textwidth]{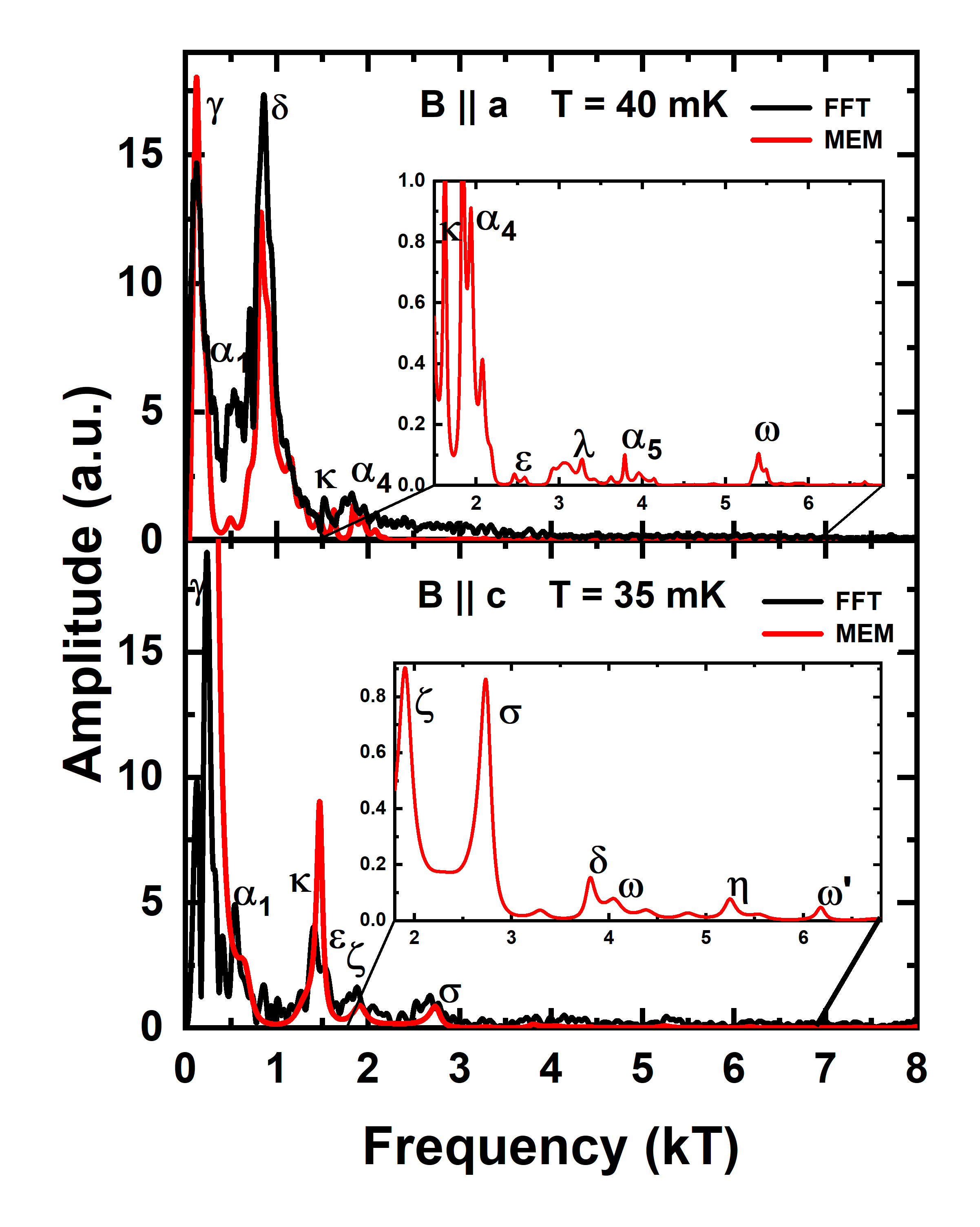}
\vspace{2cm}
\caption{\label{fig2}The FFT (black) and MEM (red) analysis are shown for B $<$ B$_m$ $||$ a (top) and  for B $<$ 24.8 T $||$ c (bottom). The insets show a zoom in for the MEM at higher frequencies, where clear peaks are present.  These peaks are consistently and systematically reproduced for all angles we investigated between c-axis and a-axis as well as for the a-b plane.}
\end{figure}

\begin{figure}[h]
\includegraphics[width=0.49\textwidth]{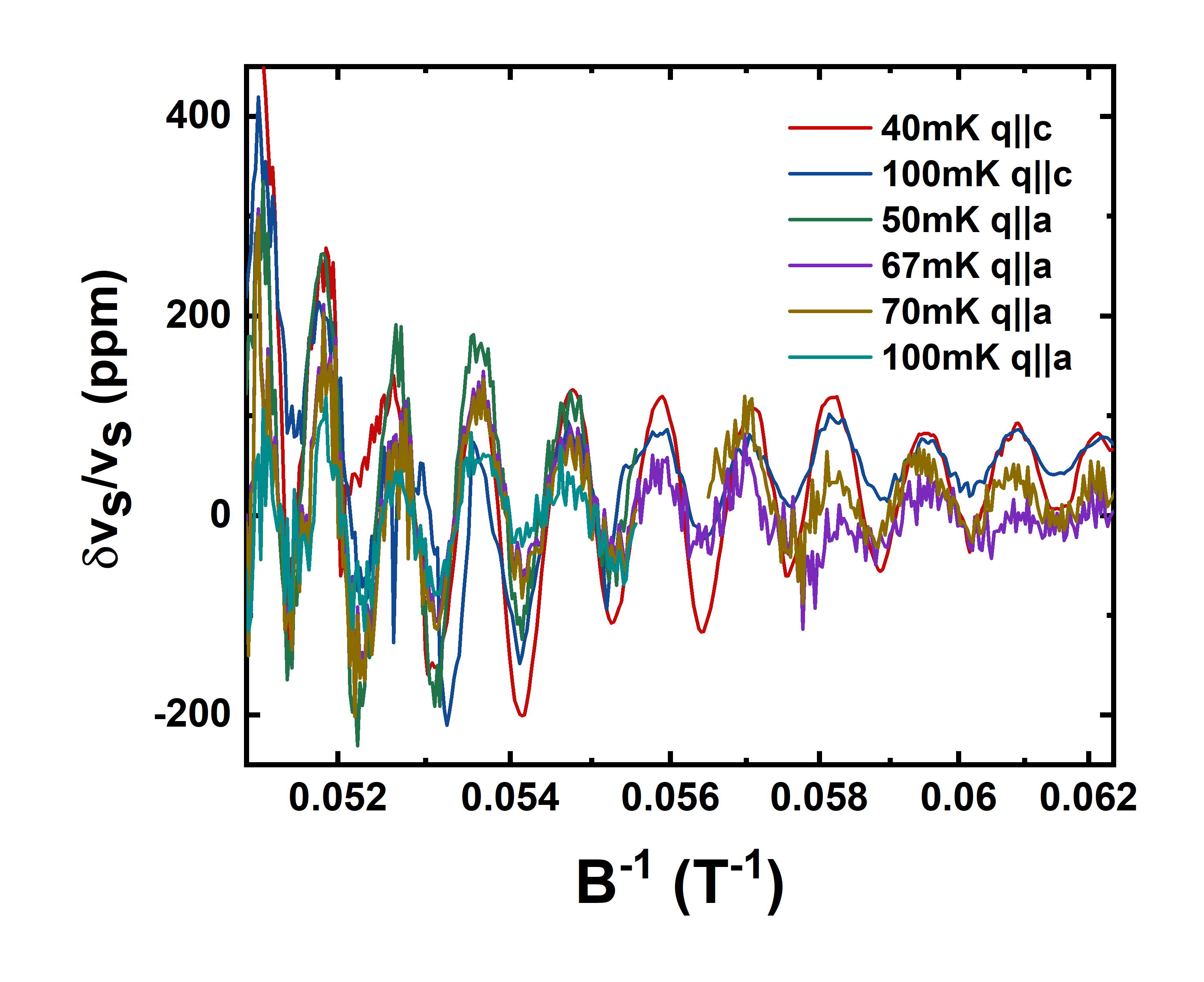}
\vspace{2cm}
\caption{\label{fig3}Shows the consistency of the MAQO frequency though the direction of propagation of the sound wave has two orthogonal orientations with respect to the magnetic field, B, which is held fixed with respect to the crystal axes.}
\end{figure}

\begin{figure*}[t]
\includegraphics[width=0.99\textwidth]{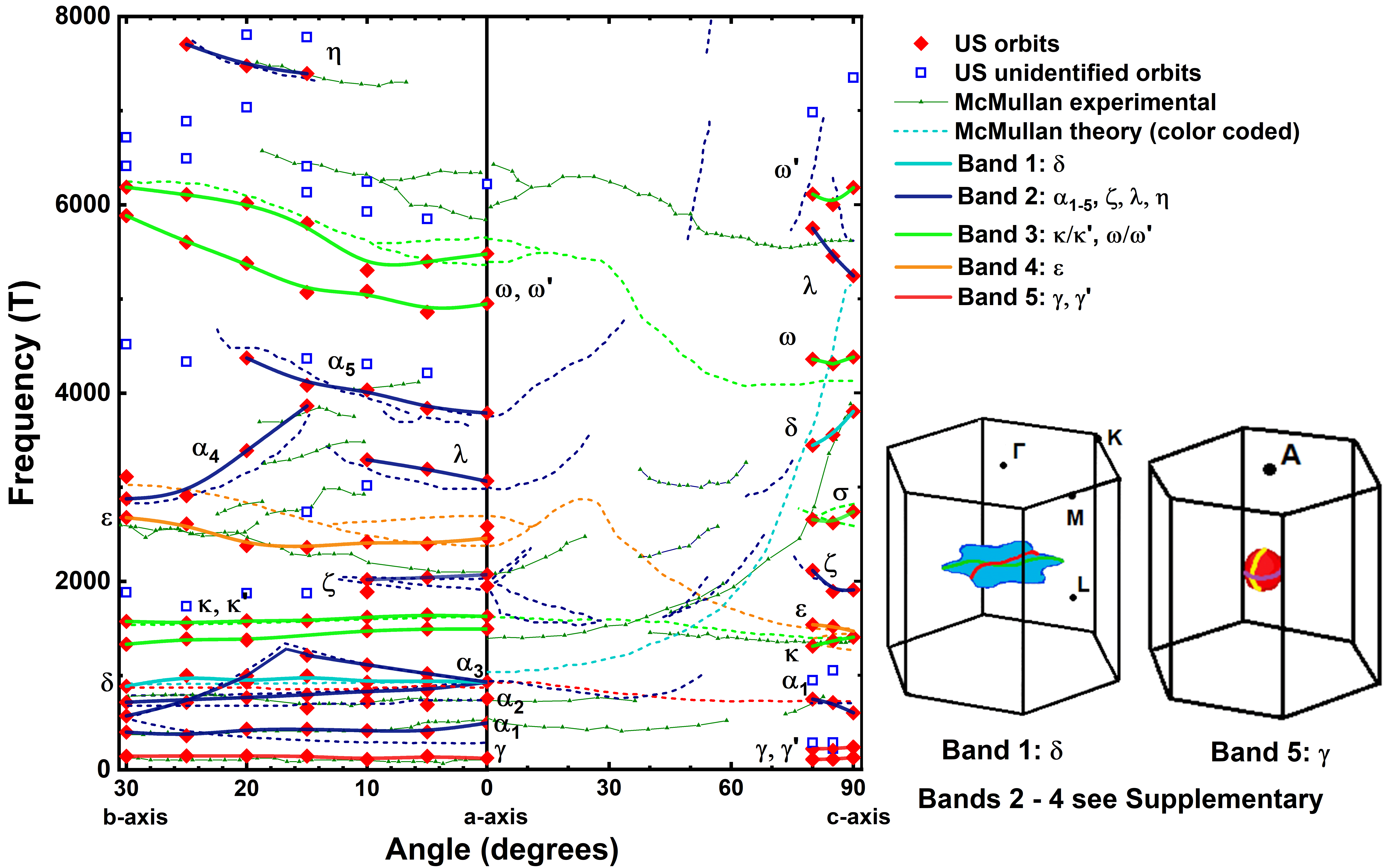}
\vspace{2.2cm}
\caption{\label{fig4}MAQO frequencies are shown vs angle for the ab-plane and a few angles from the c-axis. Red diamonds are our US-data with corresponding theoretical orbits (black lines \cite{McMullan2008}), open blue squares are peaks not previously observed in either theory or experiment. Blue triangles and lines are from magnetoresistance by McMullan\cite{McMullan2008}. The different orbits are color coded as shown on the right and match the color of the FS shown on the right and in fig. S1. Note that all subsequent plots follow the same color code for the ease of comparison.}
\end{figure*}

\textbf{IV. Analysis:} In order to analyse the MAQO, Fast-Fourier transformations (FFTs) were performed on the background subtracted data \cite{Note2}. Irrespective of the orientation of the field the MAQO emerge only around $\approx$ 12 T and are observable without a break along the a-axis up to 20 T where the MM transition interrupts the oscillations. Similarly along the c-axis the continuously observable field range is limited by the occurance of the LT at 24.8 T. In addition, as stated earlier in an ultrasonic experiment smaller orbits on the Fermi surface tend to be emphasized. The long periods together with the limited field interval for the FFT results in the resolution of the FFT being comparitively limited. To circumvent this problem, we use the maximum entropy method (MEM) \cite{Press1988} which has been shown to be a powerful tool for the analysis of quantum oscillations \cite{Sigfusson1992}. MEM analysis has a much higher frequency resolution than FFT in small intervals as it does not rely on assumptions on the oscillation behaviour outside the data range \cite{Sigfusson1992}. This allows us to make statements about the high frequency end of the spectrum as well. The results of the conventional FFT analysis as well as the result from the MEM analysis for B$\parallel$a-axis as well as the c-axis for low fields are shown in fig. 2. Indeed, in the MEM analysis distinct peaks appear in the frequency spectrum up to 8 kT despite the small amplitudes from orbits at these frequencies. Unfortunately, MEM has a downside. While the frequency value of the peaks are very precise, the amplitude values of the peaks maybe difficult to interpret \cite{Sigfusson1992}. Furthermore, as shown by Terashima et al., MEM does not provide the amplitude but rather its squared value \cite{Terashima2016}. Hence it may not be used to estimate effective masses in a straightforward manner \cite{Aoki2016, Terashima2016}.  Nevertheless, with the conventional FFT analysis we consistently see that the amplitudes at the low frequency end is always higher than the ones at higher frequencies.  This is indeed to be expected since smaller orbits produce larger amplitudes in MAQO. More quantitatively one can write the equation for the oscillatory part of the sound velocity, $\tilde{v}$, as \cite{Shoenberg1984}:\\

${|\tilde{v}| \over v}={1 \over 2} {\tilde {c}_{ikpq} \over c_{ikpq}} = {-1 \over 2c_{ikpq}} \sum_r   ({\partial lnF_r \over \partial \varepsilon_{ik}})    ({\partial lnF_r \over \partial \varepsilon_{pq}}) {d\tilde{M} \over dH} H^2$  ..(Eq.1)\\

\begin{figure}[h]
\includegraphics[width=0.49\textwidth]{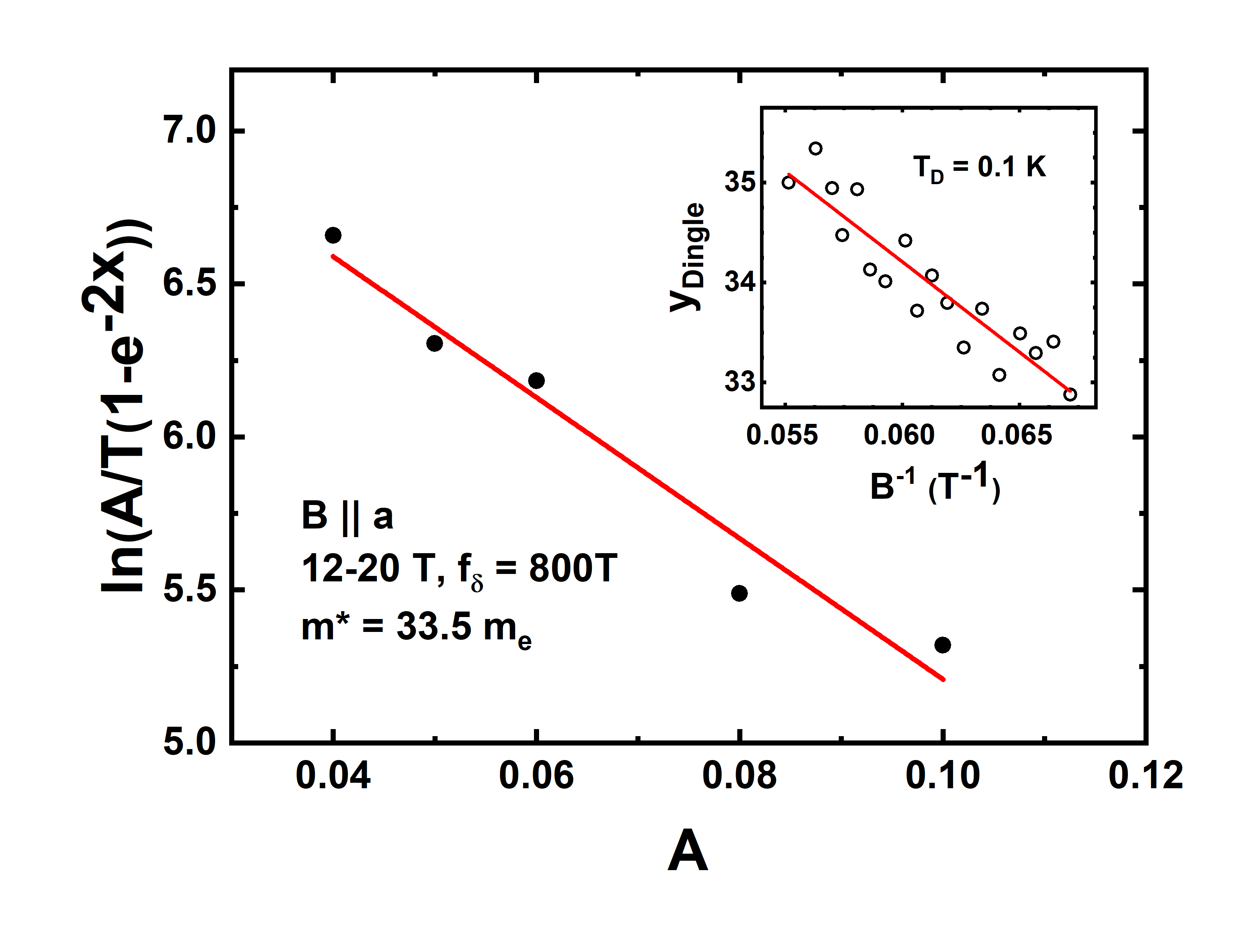}
\vspace{2cm}
\caption{\label{fig5}Effective mass plot for f$_\delta$ = 800T and B$\parallel$a in the low field regime. Amplitudes were determined from raw data with x = $\frac{2 \pi^2 m^* c k_B T}{e \hbar H}$. A least square fit gives an effective mass of m$_\delta$ = 33.5 m$_e$. The inset shows a Dingle plot from the field dependence of the amplitude. y$_{Dingle}$ = $ln(A_P H^n sinh(\frac{\alpha p T}{H}))$ where p is the pth harmonic, $\alpha$ = 1.469 10$^5$ $\frac{m}{m_e}$ $\frac{Gauss}{K}$ and n = 5/2 \cite{Shoenberg1984}.}
\end{figure}

\begin{figure}[b]
\includegraphics[width=0.49\textwidth]{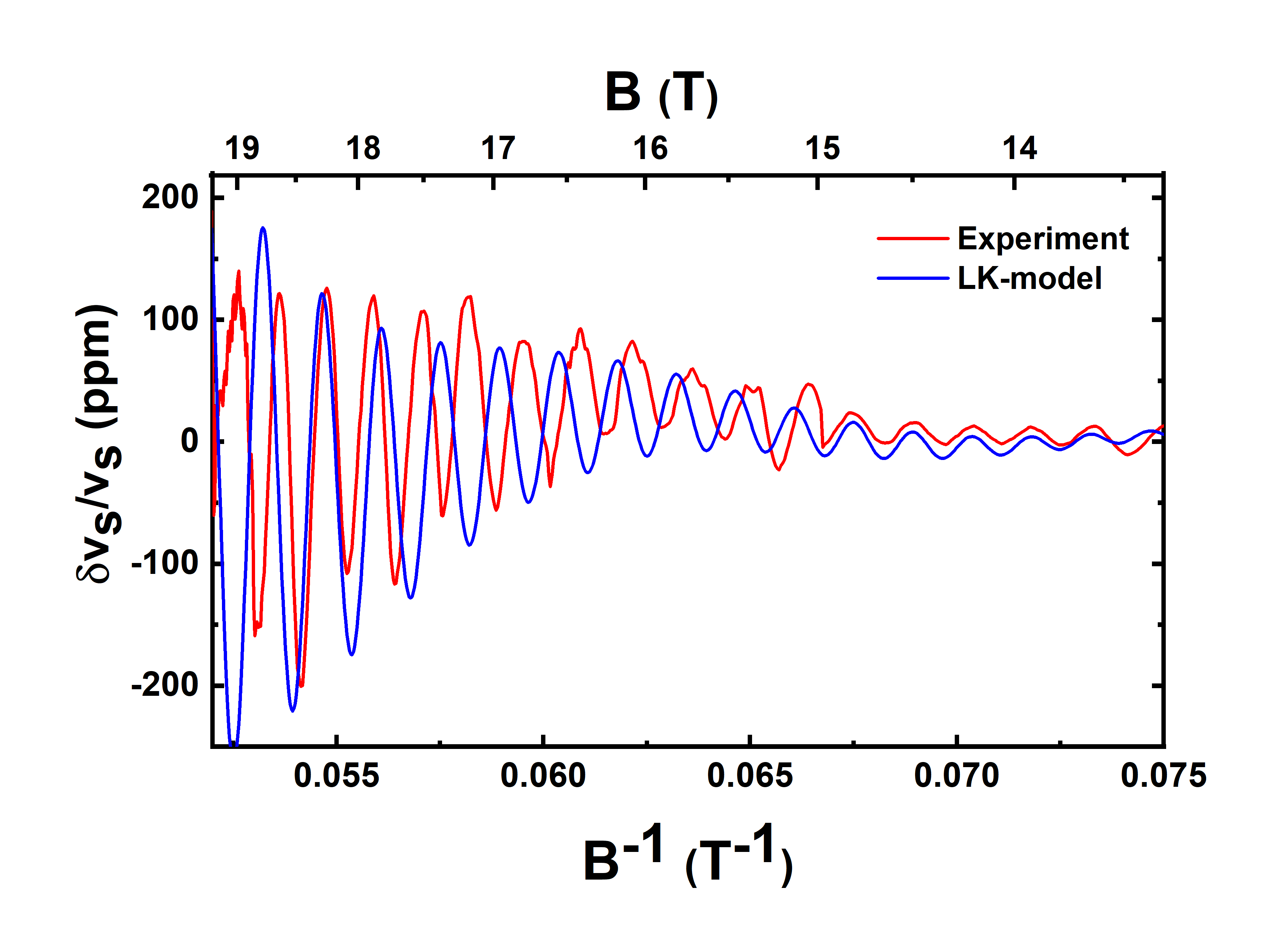}
\vspace{2cm}
\caption{\label{fig6}Shows the modeling of the MAQO using only two orbits in the equation for the sound velocity for the case of B$\parallel$a-axis.  While the model fits the data on the low field end satisfactorily it develops significant deviations in frequency at higher fields.}
\end{figure}

where the strain dependence enters clearly through the factors $(\partial lnF/\partial \varepsilon)$ with F being the MAQO frequency and the summation is across various orbits with index r.  The  $c_{ikpq}$ are the components of the elastic modulus and $\varepsilon_{ik}$, $\varepsilon_{pq}$ are the associated strains.  $\tilde{M}$ is the oscillatory part of the magnetization as per the Lifshitz-Kosevich equation, it being the first derivative of the free energy and the sound velocity being the second derivative. It is significant that despite the Gruneisen parameter for UPt$_3$ being nearly two orders of magnitude larger than for ordinary metals the maximum amplitude of the MAQO seen here is roughly the same order as those observed for the nobel metal necks. While the amplitude of the MAQO might depend on the specific strain wave (elastic mode) being employed for the measurement, the observed frequency, which is related to the extremal area of the FS through the Onsager relation, must nevertheless be the same.  We illustrate this in fig. 3 which shows measurements obtained for the same field direction but two orthogonal orientations of the wavevector, q.\\

\textbf{V. Comparison with Band Structure}: The orbital frequencies obtained using both FFT and MEM analysis for various angles of the field (B less than 20 T) in the ab-plane as well as for three different angles close to the c-axis are shown in fig.4 together with previous theoretical and experimental work\cite{McMullan2008}. In presenting our results we will follow the labeling of the orbits and the numbering of the different FS bands as presented by McMullan et al. In fig. 4 we have removed data points of higher harmonics (and single data points) to highlight the development of orbits with angle. The orbits $\epsilon$ (Suppl. Inf. fig. S1  band 4) and $\delta$ (fig. 4  band 1) observed previously in dHvA \cite{Kimura2000} and SdH \cite{Rourke2012} measurements and computed from band structure can be identified in the MAQO thus giving validation to our measurements. The same holds true for the  large $\omega$ orbit (Suppl. Inf. fig. S1 band 3), with the frequency being slightly lower than previous experiments. Due to the high sensitivity of ultrasound to small orbits, new orbits, namely $\kappa$, $\alpha_3$ and $\xi$, predicted by theory \cite{Rourke2012} are seen below 3 kT. 

The $\kappa$ orbit of electron band 3 in green, previously only seen from the c-axis to 45$^0$ off c-axis, can be followed in the ab-plane, confirming the theoretical calculation of a spherical FS (Suppl. Inf. fig. S1 band3) centered at the H or K point \cite{Rourke2012}. Both orbits of band 3, $\kappa$ and $\omega$ are observed twice at close frequencies to each other. This suggests that band 3 (see supplementary information for illustration) is particularly sensitive to spin splitting of the orbits. The smallest "pearl"-like FS labelled $\gamma$ is observed at the lowest frequency of $\approx$ 0.1 kT in the ab-plane and along the c-axis, a factor of about 10 smaller than predicted by theory.  Furthermore, we observe several theoretical orbits of band 2, which corresponds to a large hole FS. The orbits $\alpha_{2-4}$  as well as the orbits $\lambda$ and $\eta$ are observed almost precisely at the predicted theoretical values (fig. S1 band 2).  Only $\eta$ has been reported with this accuracy before, the other orbits were not seen previously. In addition, another flat band exists between 0.4-0.5 kT also seen in SdH measurements \cite{McMullan2008}. In theoretical calculations this orbit identified as $\alpha_1$ shows a decline going from the a-axis to b-axis which means that the "arm of the Octupus" like FS of band 2 is elliptical. As this band is flat, a circular arm of the FS is more realistic. At the c-axis, the orbits $\xi$ and $\lambda$ of band 2 are seen for the first time.\\

Large, quantum oscillations at a single frequency allow for modeling the data using the Lifshitz-Kosevich (LK) equation\cite{Shoenberg1984}. First, we want to utilize this for the calculation of the effective mass from the temperature dependence of the raw data without utilizing FFT. We do this by taking the "instantaneous" amplitudes (fig. S3) and averaging y$_{meff}$ for different H-values to obtain one point for each temperature in the effective mass plot. y$_{meff}$ is the expression on the vertical axis of fig. 5 at a mean magnetic field $B_{mean}=(B_{min}+B_{max})/2$, where B$_{min}$ and B$_{max}$ are low and high field boundaries of the chosen range\cite{Brasse2013}. The effective mass is then calculated by a self-consistent linear fit to the different tempteratures. The mass plot shown in fig. 6 for the $\delta$-orbit yields m$_\delta \approx$ 33.5 $\pm$ 1.5 m$_e$. The Dingle temperature is also calculated (as shown in inset). T$_D$ = 0.1 $\pm$ 0.01 K confirm the high quality of the sample.  Effective masses from the other orbits calculated from FFT are in the range of 8 - 50 m$_e$ as expected in heavy fermions.  \\

\textbf{VI. MAQO and High Field Effects:} While the above discussion pertains to the low field orbital frequencies a spectral analysis of the results in high fields provides further valuable results. We begin first with a discussion of the behavior of the quantum osccillations near and beyond the MM transition. Seen in the raw data in fig. 1, bottom left panel, are large MAQO below the MM transition with the associated frequency  of $\approx$ 0.8 kT as seen in fig. 7. Two main arguments help us to relate this frequency to one particular sheet of the FS.  First, considering the orbits in band 2 we note that several, namely, $\alpha_2$, $\alpha_3$ are located close by.  If band 2 were to be primary driver of the MMT in UPt$_3$, then all these orbits would exhibit large MAQO amplitudes. Since we observe that only one frequency dominates the MAQO, the $\alpha$ orbits can be ruled out.  Second, we note that the $\delta$ orbit, fig. 4, has the largest amplitude in dHvA experiments \cite{Kimura2000}. As oscillations in the US are also proportional to the dHvA amplitudes (see eq.1),  the large MAQO correspond to the $\delta$ orbit. \\

\begin{figure}
\includegraphics[width=0.49\textwidth]{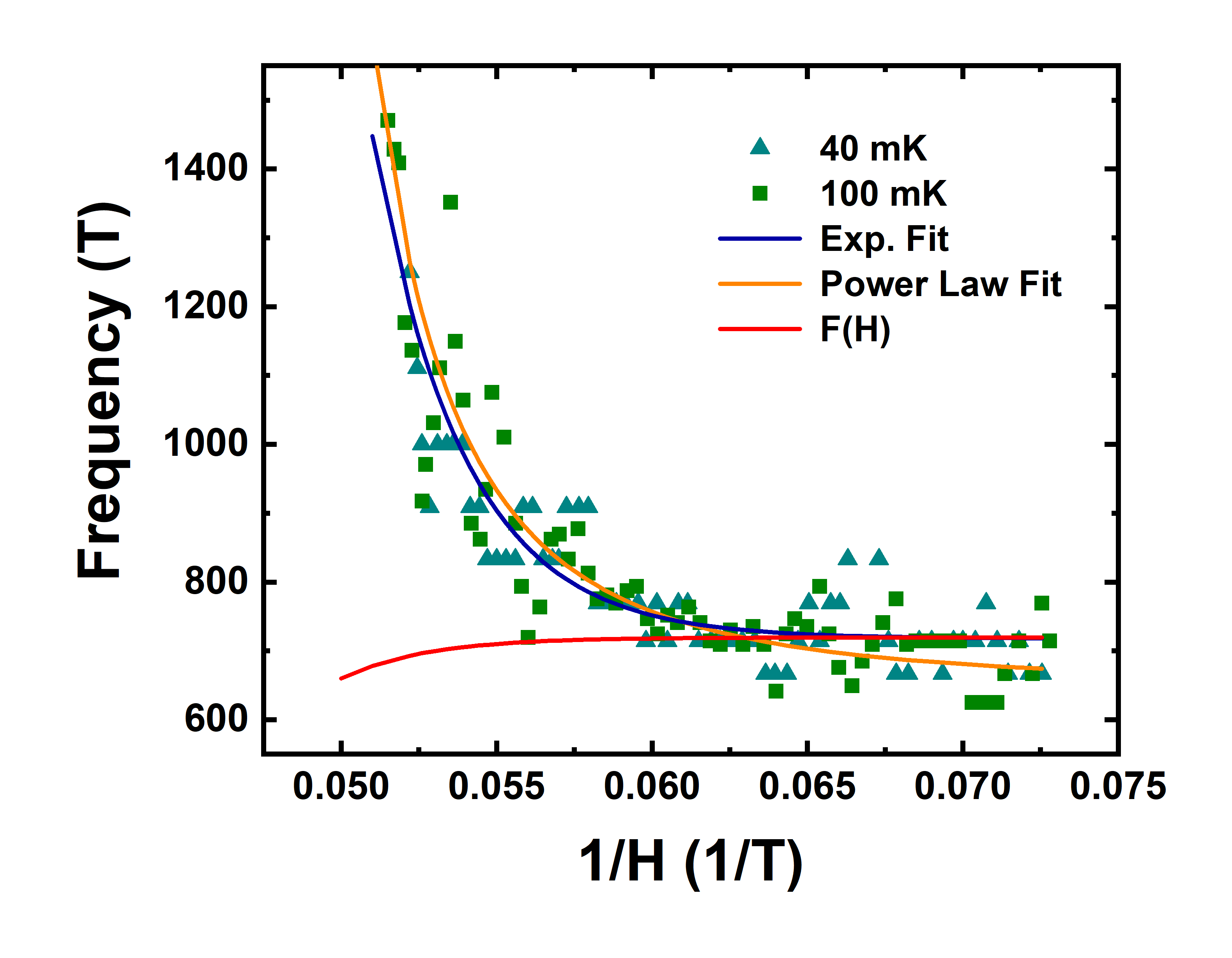}
\vspace{2cm}
\caption{\label{fig7}Shows the field dependence of the instantaneous frequency of the $\delta$ orbit which starts at 800 T at low field. The frequency increases rapidly towards the MM transition at B = 20 T. The instantaneous frequency is determined by taking the spacing between successive minima and successive maxima of the oscillations (ref. fig. 6). The red line models the field dependence as explained in the text.}
\end{figure}

Next, we try  to model the large MAQO close to the MM transition using the strain dependent LK-formula for ultrasound. While we are able to reproduce the field dependent amplitude well, the modeling of the frequency which is in conformity with the data at the low field end (12 T - 15 T), quickly goes out of phase as the MMT is approached as seen in fig. 6.  The non-LK behaviour becomes apparent as a near exponential dependence of the frequency with increasing field and this is shown in fig. 7.  Generally, increases in frequency of this magnitude are abrupt  and assigned to magnetic breakdown involving the merging of two FS \cite{Shoenberg1984}. Since the increase here is more gradual than sudden, magnetic breakdown can be ruled out. Such field dependence of quantum oscillations have been observed in a number of experiments before \cite{Ruitenbeek1982, Aoki1994} and its interpretation can be convulated. A simple explanation may be derived within a simple model for a paramagnet close to a magnetic instability.  Due to the non-negligible exchange interactions a significant splitting arises between the spin up and spin down bands and gives rise to the relation F$_{ob}$ = F(H) - H$ \partial$F(H)/$\partial$H, where F$_{ob}$ is the measured (instantaneous) frequency and F(H) is the frequency which yields the true orbital area as per the Onsager relation \cite{Ruitenbeek1982}.  Assuming a certain functional form for F$_{ob}$ the above differential equation can be solved to obtain F(H).  We have carried out this procedure by assuming an exponential behavior of F$_{ob}$ (blue line in fig. 7) which yields the solid red line for F(H) which in fact has the opposite curvature as pointed out by Kimura \cite{KimuraPhysicaB2000}. Apart from an exponential other reasonable forms such as a power law fit to the observed frequencies yield similar results. Nevertheless, in reality the situation maybe more complex:(a) the simple linear Zeeman band splitting may not be valid, (b) the effective mass may also be field dependent and (c) the FS might in fact also alter with field. Significant increases in the effective masses of orbits towards the MM transition are quite possible.  Indeed, the linear term in the heat capacity is observed to exhibit a rapid rise in UPt$_3$ and in many other MM systems such as Sr$_3$Ru$_2$O$_7$ \cite{Borzi2004} and CeRu$_2$Si$_2$ \cite{Korbel1995, Aoki1994} etc. close to the critical field.  Thus, it is hard to arrive at definitive conclusions about the more significant situation (c) for continuous changes of the FS prior to the MMT.

However, the strong connection of the $\delta$ orbit and the MMT can be ascertained in many ways. In UPt$_3$ the MM transition shifts to higher fields as $\frac{1}{cos\theta}$ with $\theta$ the angle between the ab-plane and the c-axis\cite{SuslovIntJModPhys2002}. Remarkably, band structure calculations\cite{McMullan2008} show a similar,
almost $\frac{1}{cos\theta}$ dependence for the $\delta$ orbit going from the a- to c-axis. This orbit also dominates the dHvA (i.e. magnetization) measurements, to the extent that even several distinct higher harmonics are observed\cite{Kimura2000}. If the rapid rise in the magnetization at the MMT is due to band 1 then it is natural that this sheet of the FS contribute the largest in dHvA measurements where field modulation techniques are employed. The strong dispersion also explains why we observe the $\delta$ orbit along the c-axis at high fields as seen in fig.8. The MMT is not crossed here\cite{ShivaramSR2018} and hence the orbit survives in the highest fields. Therefore, the disappearance of this sheet of the FS is the primary cause of the metamagnetism in UPt$_3$.\\

\begin{figure}
\includegraphics[width=0.49\textwidth]{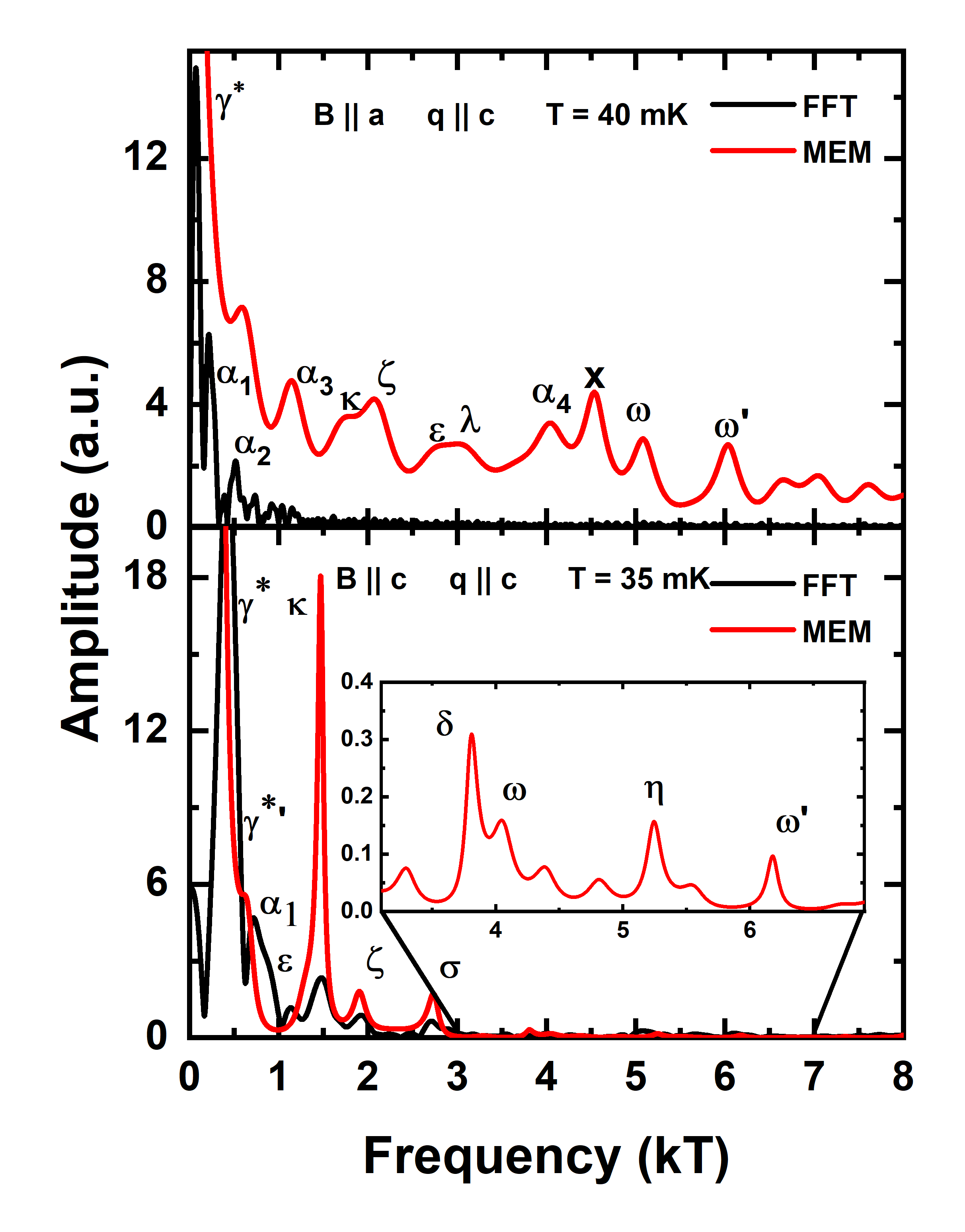}
\vspace{2cm}
\caption{\label{fig8}The FFT (black) and MEM (red) analysis are shown for B $>$ B$_m$ $||$ a (top) and  for 24.8 $<$ B $<$ 30 T $||$ c (bottom). The inset shows a zoom-in for the MEM at higher frequencies, where clear peaks are present. The peaked marked 'x' is an isolated point which did not appear at other angles.}
\end{figure}

\begin{figure}
\includegraphics[width=0.49\textwidth]{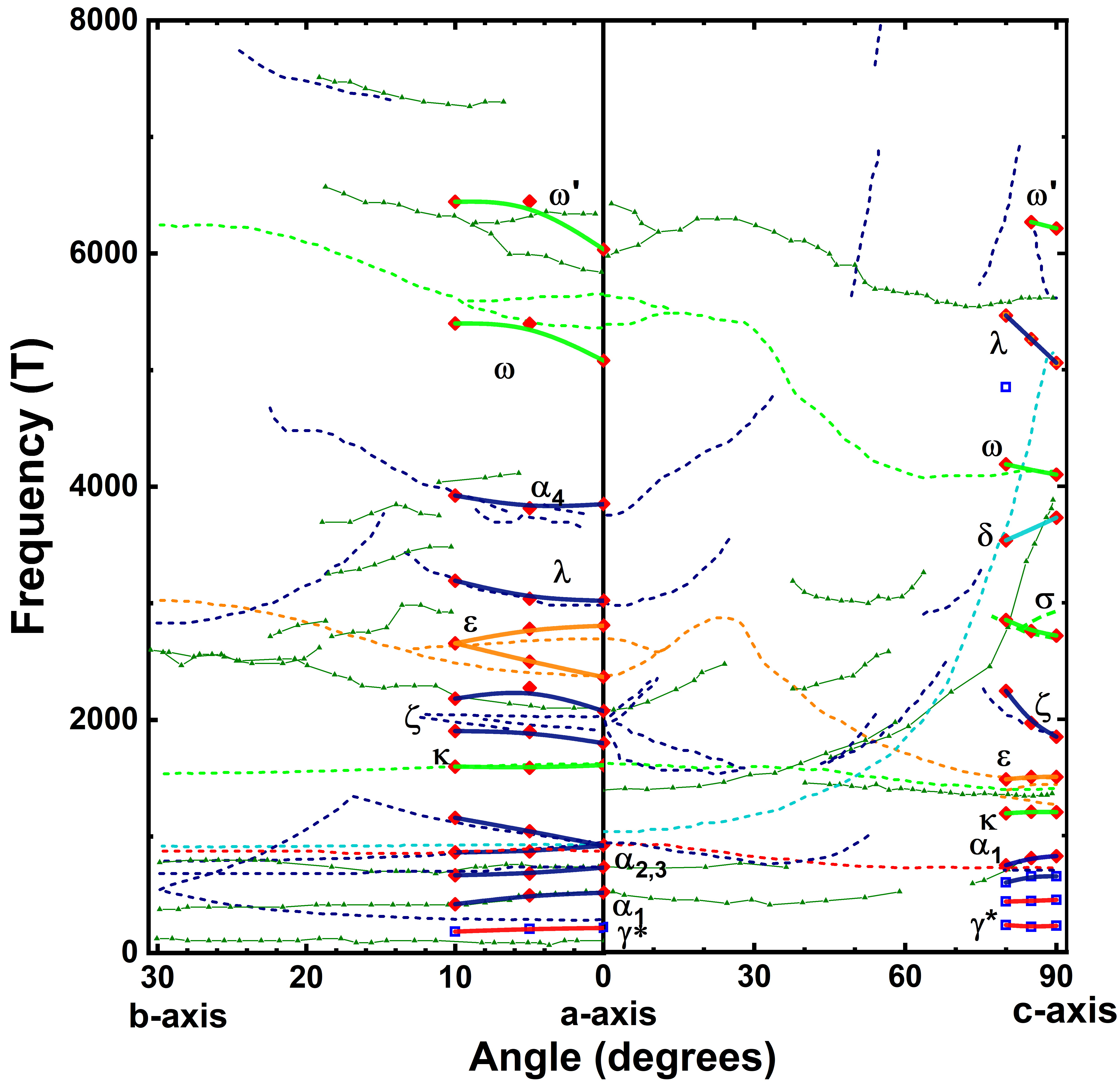}
\vspace{2cm}
\caption{\label{fig8}MAQO frequencies from FFT and MEM analysis at high fields B $>$ B$_m$ = 20T along the a-axis and B $>$ B$_{LT}$ = 24.8T along the c-axis. The color code of the data and the orbits follows figure 3. The new low frequency $\gamma^*$ orbits - which correspond to the low field $\gamma$ orbits are shown by the red lines.}
\end{figure}

In this scenario for MM in UPt$_3$ the Zeemann splitting between the spin up and spin down quasiparticle bands causes one spin band to drop below the Fermi level and disappear for higher fields. The disappearance of a whole band is thus a LT by definition\cite{Lifshitz1960}. This band associated with the large MAQO in UPt$_3$ is band 1 which disappears above B$_M$. This is in contrast to alternative FS reconstruction scenarios of the type seen in many HF compounds where a Landau type magnetic transition is involved\cite{Shishido2005, Jiao2015, Prokhnenko2017}. The large jump in the magnetization at the metamagnetic transition could very well be a consequence of this band rearrangement where instead of local moment ordering a type of itinerant SDW transition ensues. The precise nature of this "magnetic" transition. And indeed, whether it even occurs needs to be established via neutron scattering experiments in high fields. Such experiments are possible today\cite{Prokhnenko2017}.\\

Fig. 9 displays all the orbits we observe in the high field region for different angles. Representative FFT and MEM data for B $||$ a and c are shown in fig. 8. Along the a-axis, the high field region is naturally separated by the 20T MMT. Along the c-axis the LT transition at 24.8 T marked by the onset of very large MAQO separates the low and high field regimes. The oscillations along the c-axis at high field can be used for effective mass calculation (see fig. S6). For the 450 T orbit we find
m$_e$ = 31.5 $\pm$ 3.5 m$_e$ with a Dingle temperature of 0.14 $\pm$ 0.02 K. Amplitudes from FFTs for other frequencies give effective masses up to 50 m$_e$ along the a-axis and the c-axis. Therefore, heavy fermions are still present at higher fields\cite{Terashima1997, Borzi2004}. Interestingly, the dominant frequency of 450 T in the high field data has not been seen in other experiments or in existing band structure calculations. This orbit is likely the reconstructed version of the $\gamma$ orbit of the small "Pearl"-like FS. The small size of this FS leads to large MAQO in the ultrasound data for reasons discussed above. As explained in our previous work\cite{ShivaramSR2018}, the change of this tiny pocket of the FS is responsible for the LT at 24.8 T and 30 T. Although the most significant difference between low and high field measurements is seen for the $\delta$ and $\gamma$ orbit, we observe most other orbits predicted by existing band structure calculations on both sides. These orbits along the a-axis as well as the c-axis show no significant change with field, although some of the orbits appear at a slightly different frequency.\\

\textbf{VII. Conclusion:} The development of FS under magnetic fields is particularly interesting in heavy fermion compounds, as the the Zeeman energy $\mu_B$B is comparable to the Fermi energy\cite{Bercx2012}. Field induced changes in the FS can be due to Lifshitz transitions \cite{Aoki1994, Hackl2011, Daou2006, Schlottmann2011} where entirely new parts may arise or could arise as a result of a Landau type phase transition where entirely new parts may arise or exiisting parts completely disappear. While the MM transition has been studied extensively in UPt$_3$, it's relation to FS changes has not been established. As the experimental results above clearly indicate, the FS in UPt$_3$ changes drastically with magnetic fieldwith band 1 contributing a MAQO signal at low fields, with no evidence for it at high fields, B $||$ ab - plane. Its flat, "star"-like nature gives rise to the anisotropy of the MM transition. Thus, the disapperance of a specific part of the FS is identified here and we suggest that this is the "precursor" for the sudden increase in magnetization at the MMT. Similarly, for B $||$ c-axis it is the pearl shaped FS which is primarily involved in the LT at 30 T. Since this part of the FS is much smaller we expect the magnetization change as a consequence of this LT also to be commesurately small rendering its detection very difficult. Tracking the changes in the FS and relating them to magnetization signatures as observed in this work should serve as an excellent testing ground to benchmark modern approaches to band structure with prospects to take into account strong field dependent effects\cite{KentScience2018}.\\

\textbf{Acknowledgements:} We thank with pleasure Eric Palm, Tim Murphy and Scott Hannah of the NHMFL for their kind help during these measurements.  Our thanks are also due to John Singleton, Mike Norman, Igor Mazin and Michelle Johannes for many helpful discussions.  This work was supported by NSF Grant DMR-9624468.  Additional support for the work performed at the National High Magnetic Field Laboratory was provided by NSF Cooperative Agreement No. DMR-9527035 and by the State of Florida.

\begin{figure}[t]
\end{figure}

\end{document}



\begin{center}
\textbf{Supplementary Information \\
Magneto Acoustic Quantum Oscillations in High Fields and the Fermi Surface of UPt$_3$}\\
{ Ludwig Holleis, V.W. Ulrich, and B.S. Shivaram}\\
{Department of Physics, University of Virginia, Charlottesville, VA. 22904}\\

\end{center}

This supplementary information contains a schematic of Fermi Surfaces in UPt3 adapted from McMullan et al.\cite{McMullan2008} as well additional raw data, effective mass plots and FFT and MEM calculations for different field orientations and values as well as different temperatures.

\begin{figure}[h]
\includegraphics[width=150mm]{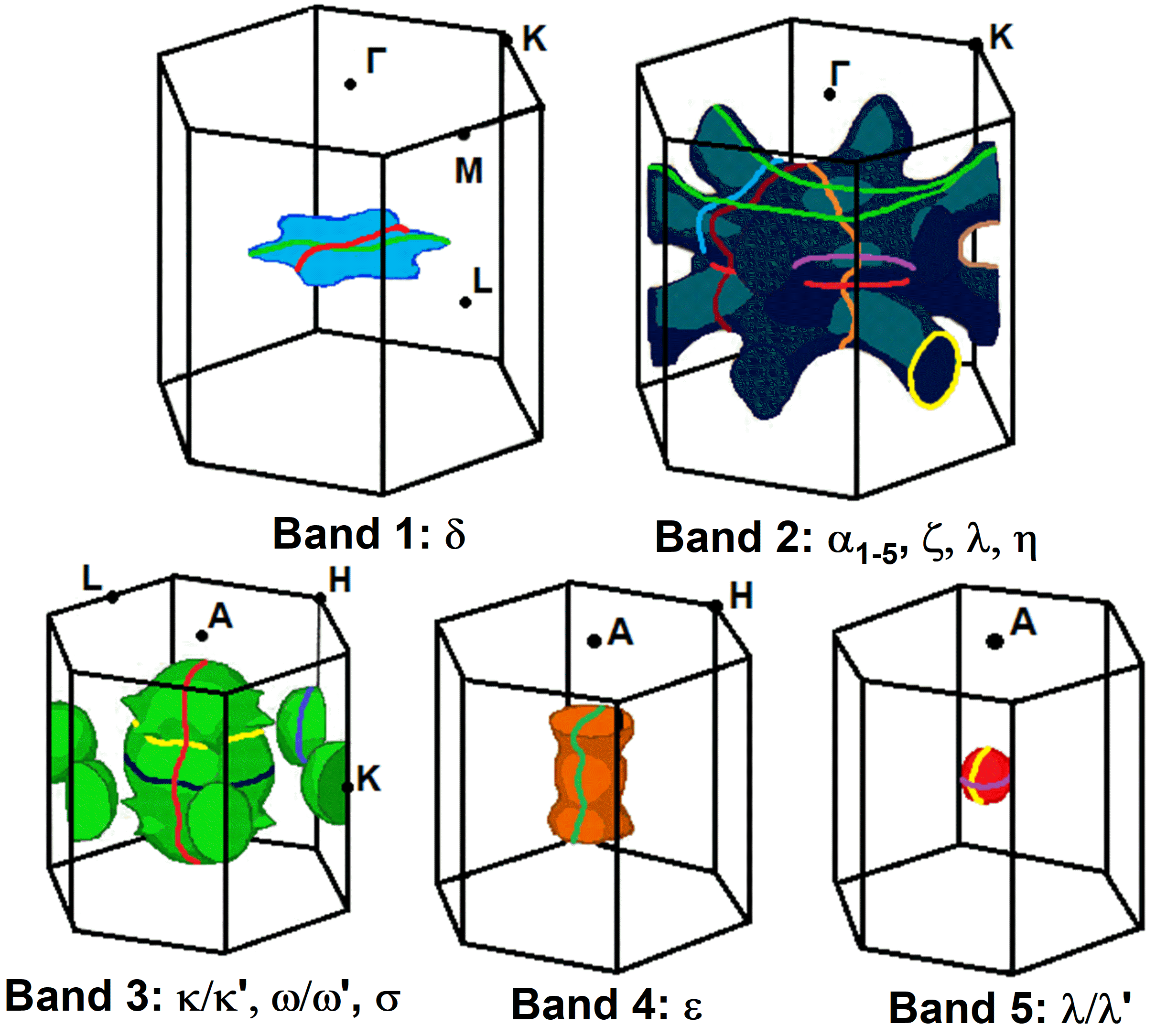}
\setlength{\abovecaptionskip}{-0pt}
\caption{Cartoon Fermi Surface with orbits reproduced from the band structure calculations of MCMullan et al.\cite{McMullan2008}. The different sheets of the FS are color coded to match the orbits in figures 4 and 9 in the text.}
\end{figure}

\begin{figure}[h]
\includegraphics[width=120mm]{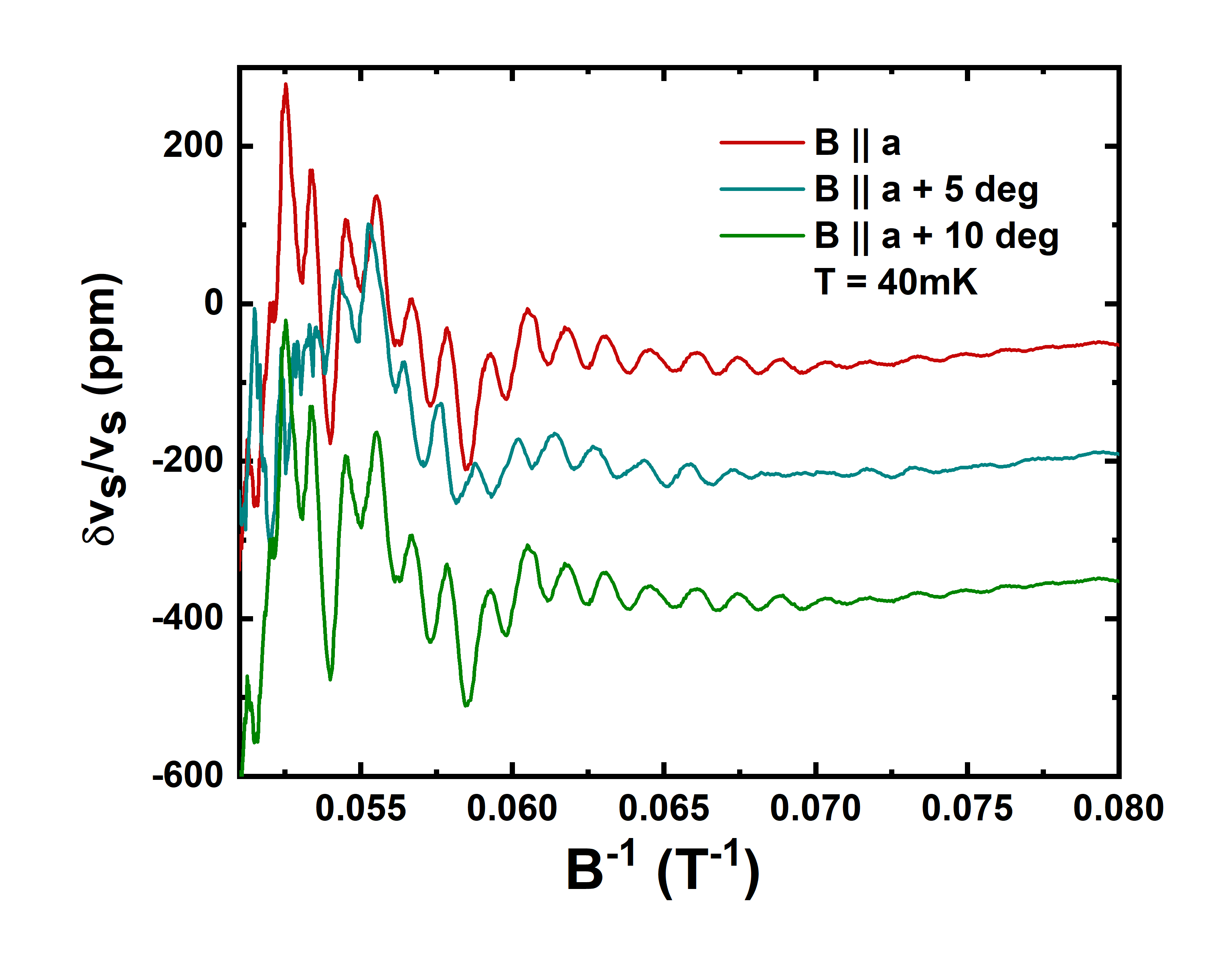}
\setlength{\abovecaptionskip}{0pt}
\caption{Raw data for B $||$ a + $\phi$ and q $||$ c at T = 40mK.}
\end{figure}

\begin{figure}[h]
\includegraphics[width=120mm]{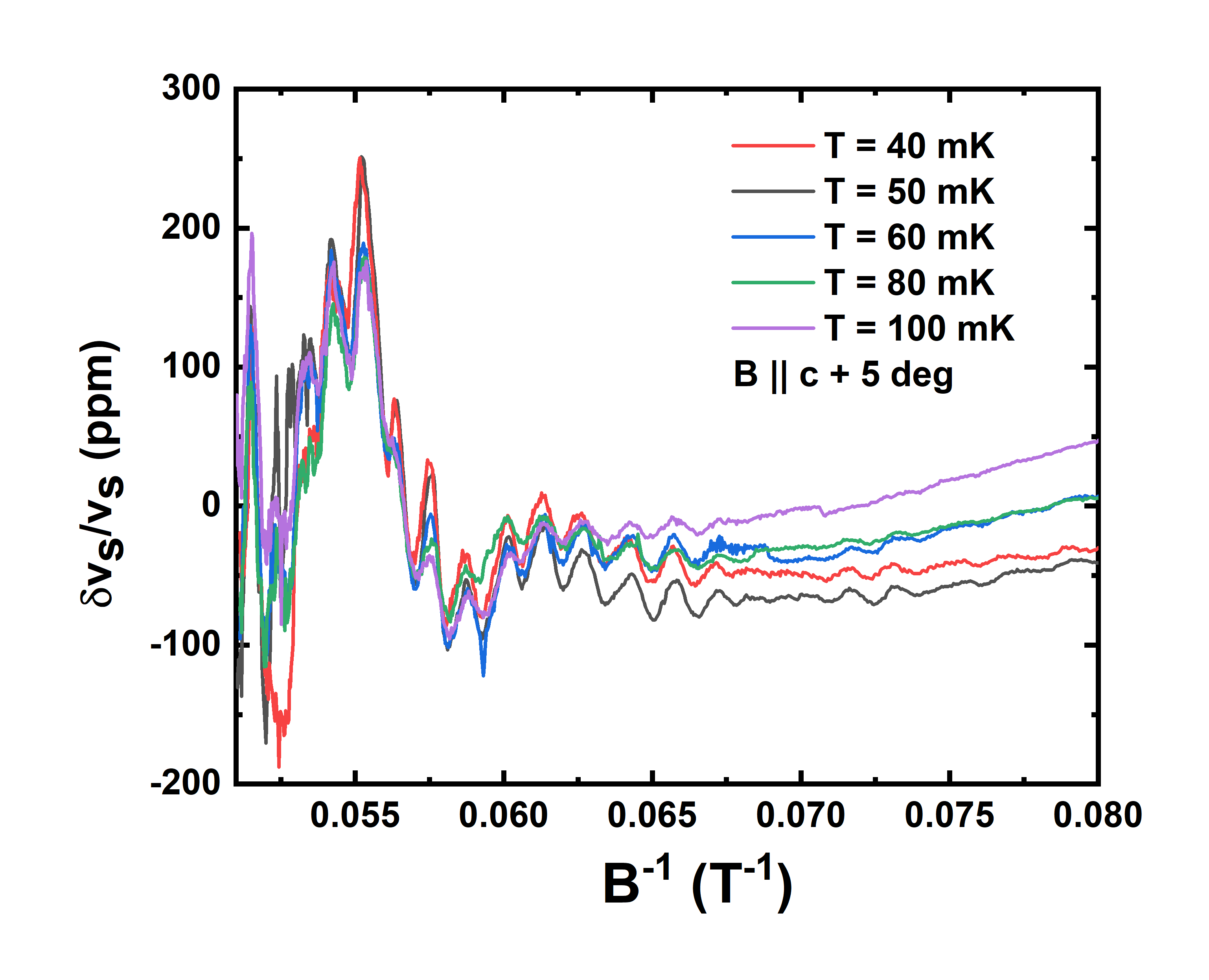}
\setlength{\abovecaptionskip}{0pt}
\caption{Raw data for B $||$ a$_{5deg}$ and q $||$ c for different temperatures. The rapid decrease in amplitude with temperature can be clearly seen, indicating a large effective mass. The corresponding effective mass plot is shown in the main text fig. 5.}
\end{figure}

\begin{figure}[h]
\includegraphics[width=120mm]{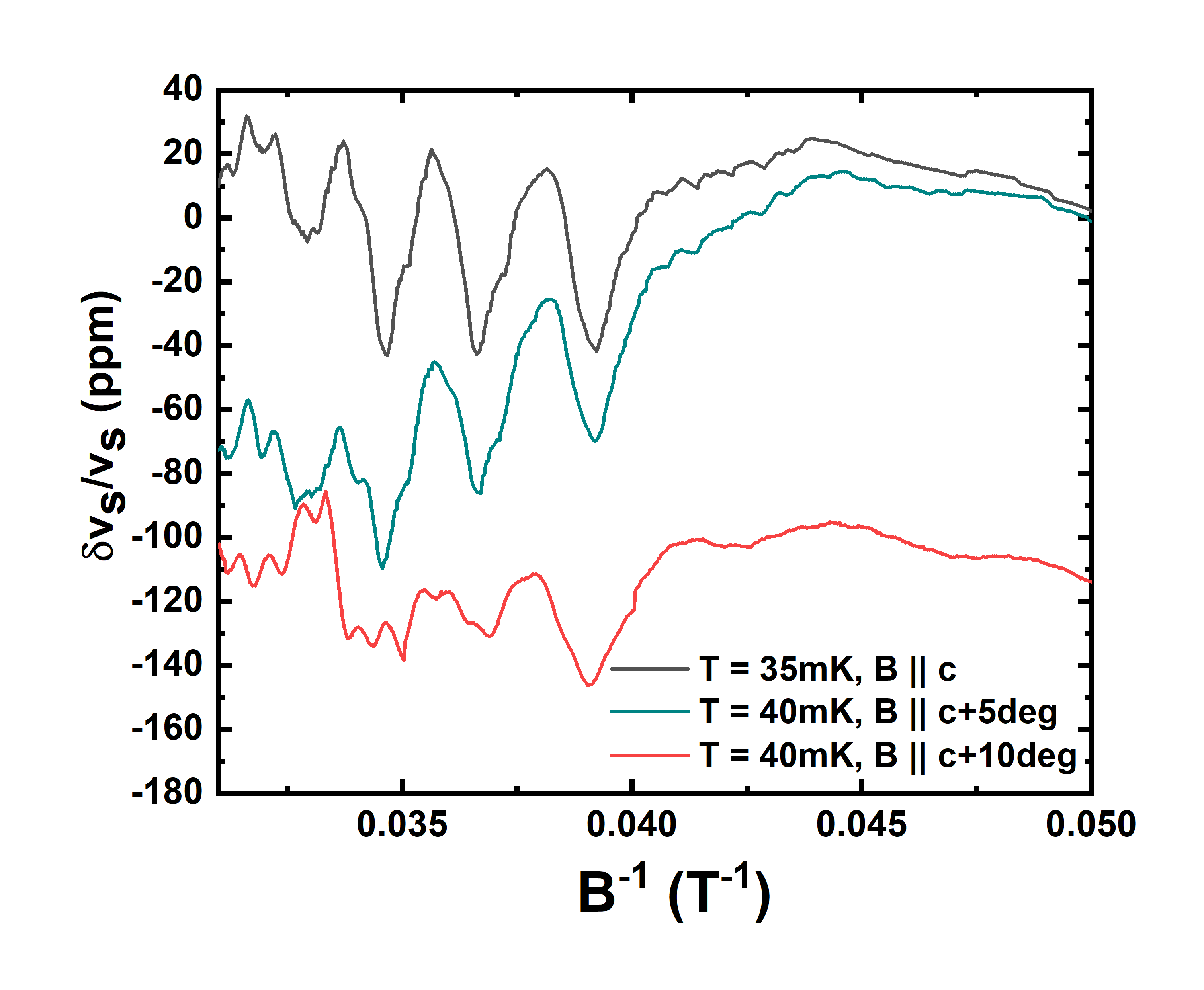}
\setlength{\abovecaptionskip}{0pt}
\caption{Raw data for B $||$ c + $\theta$ and q $||$ c at T = 35 and 40mK.}
\end{figure}

\begin{figure}[h]
\includegraphics[width=120mm]{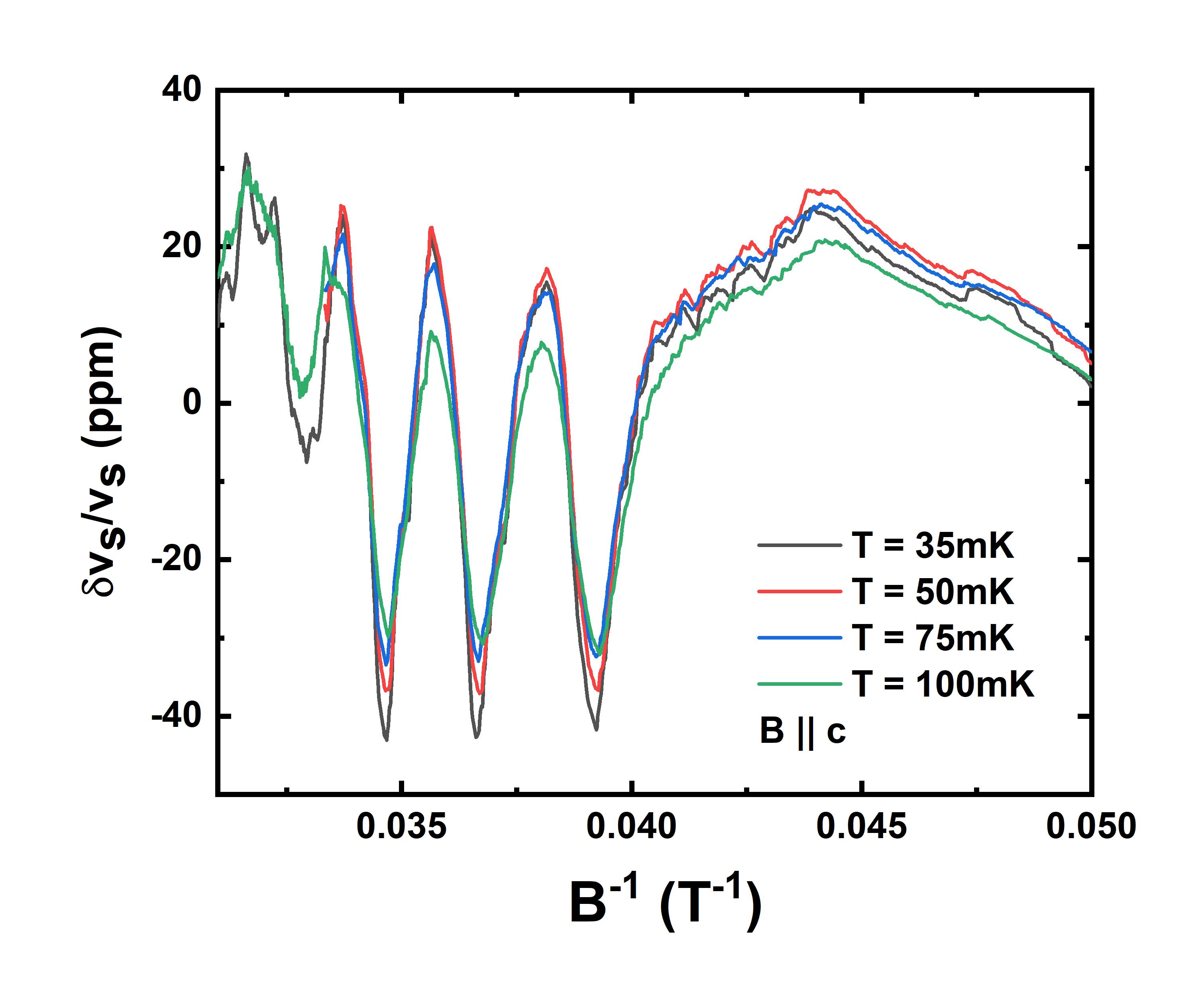}
\setlength{\abovecaptionskip}{0pt}
\caption{Raw data for B $||$ c and q $||$ c for different temperatures. The rapid decrease in amplitude with temperature can be clearly seen, indicating a large effective mass. The corresponding effective mass plot is shown in Fig. S6.}
\end{figure}

\begin{figure}[h]
\includegraphics[width=120mm]{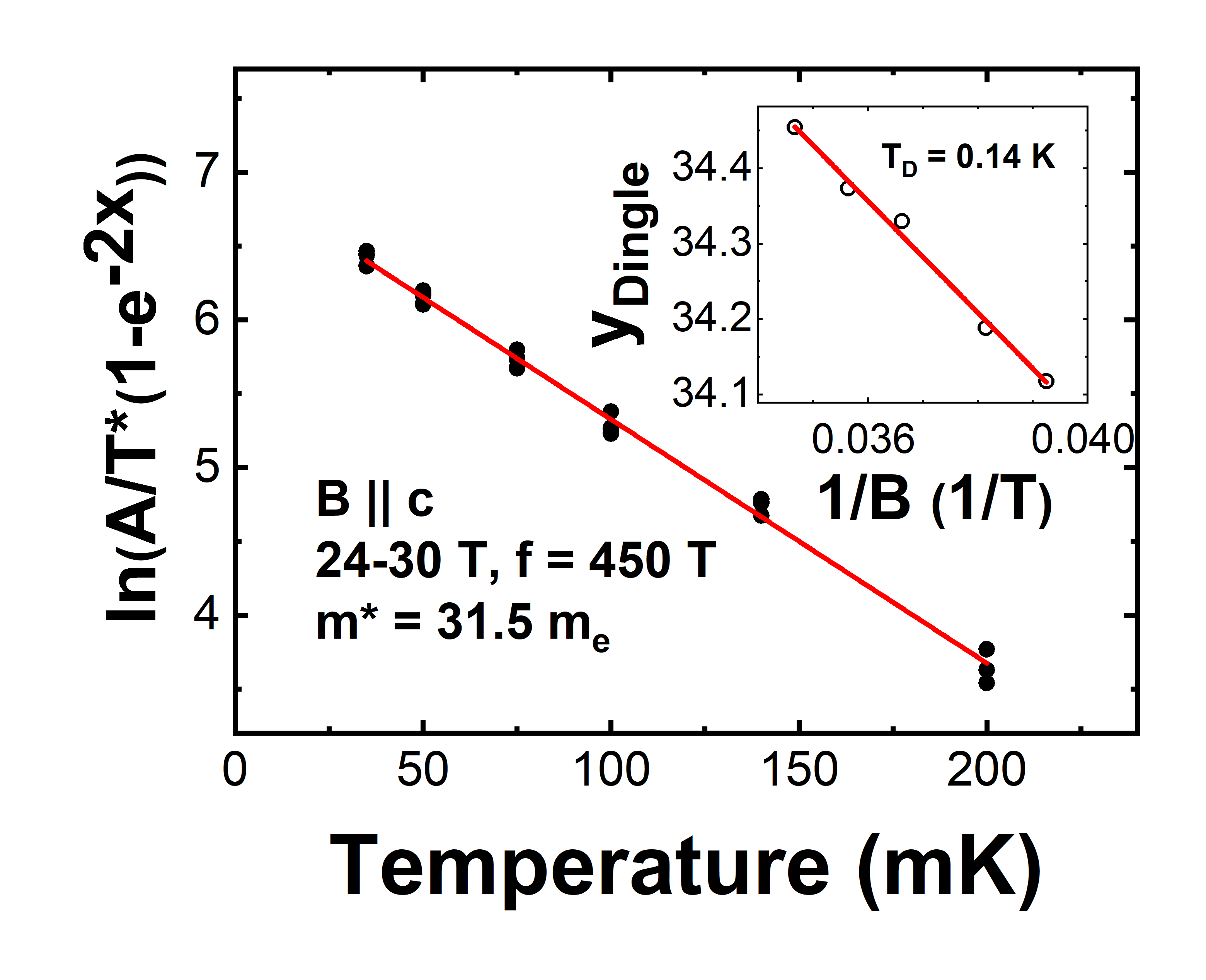}
\setlength{\abovecaptionskip}{0pt}
\caption{Effective mass and Dingle plot for the dominant MAQO of 450 T between 24.8 T and 30 T.}
\end{figure}

